\documentclass{ws-procs9x6-cpt16}
\begin{document}

\newcommand{\refeq}[1]{(\ref{#1})}
\def\etal {{\it et al.}}

\title{Calculating the Finite-Speed-of-Light Effect in\\
Atom Gravimeters with General Relativity}

\author{Yu-Jie Tan and Cheng-Gang Shao}

\address{MOE Key Laboratory of Fundamental Physical Quantities Measurement\\
School of Physics, Huazhong University of Science and Technology\\
1037 Luo Yu Road, Wuhan 430074, People's Republic of China}

\begin{abstract}
This work mainly presents a relativistic analytical calculating method for the finite speed-of-light effect
in atom gravimeters, which can simplify the derivation and give a more complete expression for the associated correction.
\end{abstract}

\bodymatter
\section{Introduction}
The finite-speed-of-light (FSL) effect has been studied by other research groups, and they
gave different results.\cite{1,4} In this article, we present an analytical relativistic study method to recalculate this effect.

\section{Starting point of the method}
In an atom-gravimeter system, the total phase
shift can be written as the sum of three components:
$\Delta {\phi _{{\rm{tot}}}} = \Delta {\phi _{{\rm{propagation}}}} + \Delta {\phi _{{\rm{laser}}}} + \Delta {\phi _{{\rm{separation}}}}$.\cite{4}
For calculating the propagation phase shift $\Delta {\phi _{{\rm{propagation}}}}$, one should
first perform integrals of the lagrangian 
along the upper and lower paths over time
to obtain the actions, and then take the difference between them. When one considers
that the speed of light is finite, the calculation is complex, since the integral intervals
for the two paths are different (see Fig.~1 left). To simplify the calculation, we propose analyzing
the system in a new coordinate system, where the integral intervals can be synchronized.
Then, taking the first pulse separation for example, the difference between the actions for
the two paths will undergo the change below:
\begin{equation}
\int_{{t_A}}^{{t_B}} {{L_{{\rm{upper}}}}} ~dt - \int_{{t_A}}^{{t_C}} {{L_{{\rm{lower}}}}} ~dt \to \int_{{t'_A}}^{{t'_N}} {({L'_{{\rm{upper}}}} - {L'_{{\rm{lower}}}})} ~dt',
\label{aba:eq1}
\end{equation}
which can save a lot of unnecessary calculation. Thus, the crucial step for our method is making a coordinate
 transformation for the laser beam.

\begin{figure}
\begin{center}
\includegraphics[width=\hsize]{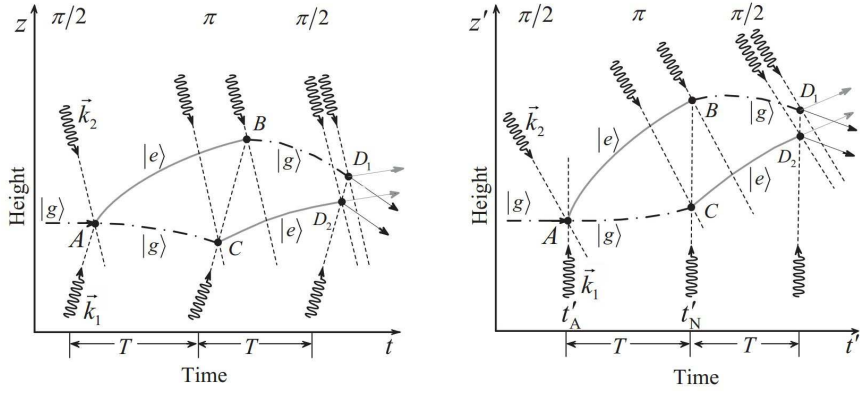}
\end{center}
\caption{The spacetime diagram of a light-pulse in an atom interferometer before (left) and after (right) the coordinate transformation.
}
\label{aba:fig1}
\end{figure}

\section{Coordinate transformation}
We assume light $\vec{k_{1}}$ is the one reflected by the mirror in the bottom of the experimental
setup. Based on the analysis in Ref.\ \refcite{4}, light $\vec{k_{1}}$ determines the change 
of the atom's state, and it can be considered as the ``control light.''

First, we should solve the geodesic equation of the photon to derive its trajectory $t=f(z)$, and then make
a coordinate transformation for light $\vec{k_{1}}$:
\begin{equation}
\left\{\begin{array}{ll}
t' = t - f(z),\\[2mm]
\vec{z}{}' =  \vec{z}.
\end{array}\right.
\label{aba:eq2}
\end{equation}
After this transformation, the coordinate velocity of $\vec{k_{1}}$ undergoes the change $c\rightarrow+\infty$
(see Fig.\ 1 right), and the lagrangian $L'$ of the atom can be written as the sum of a quadratic part and a nonquadratic part:
$L'=L'_{\rm{quad}}+L'_{\rm{nonquad}}$. Thus, the propagation phase shift can be further expressed as:
\begin{equation}
\Delta {\phi _{{\rm{propagation}}}} = \Delta {\phi _{L'_{\rm{quad}}}} +\Delta {\phi _{L'_{\rm{nonquad}}}}.
\label{aba:eq3}
\end{equation}

\section{Calculating the phase shift and the measured $g$}
For $\Delta {\phi _{L'_{\rm{quad}}}}+\Delta {\phi _{{\rm{separation}}}}$, one can calculate
 it by combining the Bord\'e ABCD matrix method with quantum mechanics, which was previously studied in Ref.\ \refcite{5}:
first derive the quadratic hamiltonian through the quadratic lagrangian, and further solve the motion equation of the
 atoms, and finally insert them into the action of the atoms (in the ABCD matrix form) to get the related phase shift.

For $\Delta {\phi _{L'_{\rm{nonquad}}}}$, one can refer to Ref.\ \refcite{6}: treat $L'_{\rm{nonquad}}$ as a
perturbation for $L'_{\rm{quad}}$, and use the perturbation approach to calculate the phase shift.

For $\Delta {\phi _{{\rm{laser}}}}$, the frequency chirps should be taken into consideration. Taking light $\vec{k_{1}}$, for
example, one can simply introduce the phase as:
$- {\omega _1}(t - z/c) - \frac{1}{2}{\alpha _1}{(t - z/c)^2}$,
where ${\alpha _1}$ is the frequency chirp for light $\vec{k_{1}}$.
Then, combining the interaction between the atoms and the Raman light field, one can obtain the phase shift introduced by
the laser beams.

Consequently, through summing the phase shifts analyzed above, one can derive the total phase shift and further the measured
acceleration due to gravity, which can be kept to some high-order terms including some general relativistic effects. Considering only
 the FSL correction, the measured $g$ can be written as:
\begin{equation}
g \approx {g_0}\left( {1 + \frac{{v(T)}}{c} + 2\frac{{v(T)}}{c}\frac{{{\alpha _1} + {\alpha _2}}}{{{\alpha _1} - {\alpha _2}}}} \right),
\label{aba:eq4}
\end{equation}
with $v(T)$ the atom velocity at the $\pi$ pulse, 
and ${\alpha _1}$ and ${\alpha _2}$ the frequency chirps for the two Raman beams.

\section{Summary}
This analytical study method can be used to calculate the relativistic effects and present an analytical derivation process. From the result,
one can separate the FSL effect and obtain a more complete expression for the FSL correction.

\section*{Acknowledgments}
This work is supported by the National Natural Science Foundation of China (Grant
No.\ 11275075).

\end{document}